\documentclass[twocolumn,showpacs,preprintnumbers,unsortedaddress,%
floatfix,amsmath,prb]{revtex4}

\usepackage{bm}
\usepackage{graphicx}%
\usepackage{calc}
\usepackage{dcolumn}%

\newcommand{\bra}{\langle}
\newcommand{\ket}{\rangle}

\begin{document}

\title{Magnetic anisotropy of vicinal (001) fcc Co films:
role of crystal splitting and structure relaxation in step-decoration
effect}

\author{M.\ Cinal}
\email{mcinal@ichf.edu.pl}

\affiliation{Institute of Physical Chemistry of the Polish Academy
of Sciences, ul. Kasprzaka 44/52, 01--224 Warszawa, Poland}

\author{A.\ Umerski}
\email{a.umerski@open.ac.uk} \affiliation{Department of Applied
Mathematics, Open University, Milton Keynes MK7 6AA, United
Kingdom}

\date{\today}

\begin{abstract}

The uniaxial in-plane magnetic anisotropy (UIP-MA) constant is
calculated for a single step on the (001) surface of fcc Co($N$)
films. The calculations are done for both an undecorated step and
the step decorated with one or more, up to 7, Cu wires. Our
objective is to explain the mechanisms by which the decoration
decreases the UIP-MA constant, which is the effect observed
experimentally for ultrathin Co films deposited on vicinal (001)
Cu surfaces and can lead to reorientation of magnetization within
the film plane. Theoretical calculations performed with a
realistic tight-binding model show that the step decoration
changes the UIP-MA constant  significantly only if the splitting
between the on-site energies of various $d$-orbitals is included
for atoms located near the step edge. The local relaxation of
atomic structure around the step is  also shown to have a
significant effect on the shift of the UIP-MA constant. The
influence of these two relevant factors is analyzed further by
examining individual contributions to the UIP-MA constant from
atoms around the step. The magnitude of the obtained UIP-MA shift agrees
well with experimental data.
It is also found  that an additional shift
due to possible charge transfer between Cu and Co atoms
is very small.

\end{abstract}


\maketitle

\section{INTRODUCTION}
\label{sec-intro}

The magnetic anisotropy (MA), determined by the dependence of energy
on the orientation of magnetization, is one of the basic properties of
magnetic systems. A non-zero MA energy arises due to two different
physical effects: long-range magnetic dipole-dipole interaction and
spin-orbit (SO) coupling. The latter interaction couples the spin of
an electron to the local electric field as the electron travels
through the system which makes the interaction sensitive to the
local atomic structure. This leads to the magnetocrystalline (MCA)
component of the MA energy. As the symmetry is reduced in nanoscopic
structures, the MCA energy (per atom) becomes larger, normally by
two orders of magnitude, than in bulk cubic crystals.
\cite{chappert86, broeder92}
 An example of such low symmetry structures is a ferromagnetic
ultrathin film deposited on  a vicinal surface which
consists of monoatomic steps separated by flat terraces.
\cite{allenspach95a,wursch97}.
Vicinal surfaces can be obtained by slightly
miscutting low-index [e.g., (001)] surfaces of substrate crystal.

In the present paper, we study theoretically MA of fcc Co films
placed on a vicinal (001) Cu surface with step edges running along
the direction $[1\overline{1}0]$.
In such systems, the magnetization $\bm M$ lies
within the Co film plane due to the negative sign of the
out-of-plane MA constant $K_2$ which determines the
energy dependence $K_2 \sin^2\theta$ on the magnetization polar
angle $\theta$; cf. Ref. \onlinecite{krams92}. With  $\bm M$ lying in-plane
(i.e., for $\theta=\pi/2$), the system energy depends on the azimuthal angle
$\phi$ which $\bm
M$ makes with the $[1\overline{1}0]$  
 axis:
\begin{equation}\label{MA-energy-gen}
E(\phi)=K_{1}/4 \sin^{2}(2\phi)+K_{\text{u}} \sin^2 \phi -M H
\cos(\phi-\phi_H).
\end{equation}
Here, the first term on the right-hand side is due to  biaxial
(fourfold) MA which arises because the directions [110] and [100]
are non-equivalent in systems with cubic crystal structure. The
cubic MA constant $K_{1}$ is positive for fcc Co/Cu(001)
films.\cite{krams94} Thus, in flat Co/Cu(001) films the
magnetization lies along one of the two equivalent directions: [110]
or $[1\overline{1}0]$ , the easy axes, while the directions [100], [010] are hard
axes.
 The existence of steps in vicinal Co/Cu(001) films introduces
 an additional, uniaxial (twofold), magnetic
anisotropy within the film plane. The directions [110] and
$[1\overline{1}0]$  seize to be equivalent: one of them remains the easy axis,
the other becomes the intermediate axis.
\cite{allenspach-apl97,allenspach-jap-review}
The sign of the uniaxial in-plane MA
(UIP-MA) constant $K_{\text{u}}$, equal to
\begin{equation}\label{Ku-dif-e}
K_{\text{u}}=E(\phi=\pi/2)-E(\phi=0),
\end{equation}
at $H=0$, determines the easy direction of magnetization: $\bm M$ is
parallel (perpendicular) to the step edges  when $K_{\text{u}}$ is
positive (negative).

 The expression (\ref{MA-energy-gen}) includes, optionally, an external
magnetic field $\bm H$ which is assumed to lie within the film plane
at the
azimuthal angle $\phi_H$ to the $[1\overline{1}0]$  
axis. Experimentally, by applying the field $\bm H$ perpendicularly
to the easy axis, one can measure the so-called shift field
$H_{\text{s}}$ and determine the UIP-MA  constant as
\begin{equation}\label{Ku-Hs}
K_{\text{u}}=H_{\text{s}} M_{\text{s}};
\end{equation}
here $M_{\text{s}}$ is the saturation magnetization.
\cite{allenspach-apl97} This relation holds as long as
$K_{\text{u}}$ is much smaller than $K_1$ which is true for vicinal
Co/Cu(001) films.\cite{allenspach-apl97}

The magnetic anisotropy of thin ferromagnetic films deposited on non-magnetic
vicinal surfaces has been the subject of  many experimental papers.
Apart from Co/Cu(001) systems,
\cite{allenspach-apl97,allenspach-jap-review,allenspach95a,allenspach95b,
allenspach96a,allenspach96b,allenspach96c,kawakami98a}
 the reported work also concerned Fe
films on vicinal Ag, Au, W, Pd, Cr (001) substrates
\cite{kawakami96,qiu98,kawakami98b,kawakami99,qiu02,
hillebrands02,hillebrands05}
and vicinal Ni/Cu(001) systems.\cite{qiu00a}
 Related experimental measurements of
UIP-MA have been performed for Co/Cu(110) ultrathin films
\cite{110-as-vicinal,bland99,qiu00b} and also for Co films obtained by
depositing Co atoms at an oblique  angle on a flat (001) Cu surface
which leads to some structural anisotropy.
\cite{poelsema01} In the experiments on  MA
in vicinal films, it was investigated how the shift field $H_{\text
s}$ depends on the ferromagnetic film thickness, the miscut angle
$\alpha$ and the miscut direction.\cite{hillebrands02} These
dependencies were found to be determined by the kind of the deposited film,
its lattice type, the choice of substrate, and, in consequence, also
by the strain present in the film. For Co/Cu(001) systems, the UIP-MA
constant $K_{\text{u}}$ has been found to depend linearly on the
miscut angle $\alpha$ which implies that $K_{\text{u}}$ is
proportional to the step density; cf. Eq. (\ref{nterrace}) below.
The experimental dependence on the Co film thickness $d$ is more
complicated: the usual breakdown: \cite{ku-explanation}
$K_{\text{u}}=K_{\text{u}}^{\text{v}}+K_{\text{u}}^{\text{s}}/d$
into the volume and surface terms is not generally valid,
presumably, due to the anisotropic relaxation of
the biaxial strain present in Co/Cu systems.\cite{allenspach96c}

Covering a vicinal ferromagnetic film with non-magnetic
adsorbates, like  Cu, Ag, O or CO, usually has a large effect  on
UIP-MA. \cite{allenspach95a,allenspach95b,qiu00b} In particular,
it has been found experimentally that even submonolayer amount of
Cu deposited on a vicinal Co film can decrease $K_{\text{u}}$ so
significantly that, in some cases, $K_{\text{u}}$ becomes negative
and the magnetization direction switches to perpendicular to the
steps.\cite{allenspach95a,allenspach95b,
allenspach96a,allenspach96b,allenspach96c, allenspach-apl97,
allenspach-jap-review}.
 This effect is usually explained by change in local MCA due to Co-Cu orbital
hybridization at Co step edges after they are decorated with Cu
atoms.\cite{allenspach-sur-sci97}
Such qualitative explanation seems plausible, however it still requires
further investigation
since few, and only partly relevant, theoretical calculations have
been done so far. In Ref. \onlinecite{smirnov96-7},
the energy of a single step on a (001) fcc Co film with
non-collinear magnetic structure  was calculated within
an {\em ab-initio} model without SO coupling included.
However, no values for the  UIP-MA constants nor  definite conclusions
about the magnetization reorientation were given there.
The relativistic full-potential
linear-augmented-plane-wave calculations,
\cite{freeman98} performed for a planar network of
parallel Co wires separated with  empty  wires (i.e., a striped monolayer),
have shown that
magnetization changes its orientation  upon filling the empty wires
with Cu. However, this structure seems too simplified to extend the
obtained results to the case of Co vicinal films. In Ref.
\onlinecite{llois02}, more realistic systems, i.e., unsupported
stepped Co and  Fe monolayers were considered using a tight-binding
(TB) model, but the effect of step decoration was not studied. The
present authors reported \cite{mc-au-surf-sci}  a TB calculation for
a monoatomic step on a 5-monolayer (ML) (001) fcc Co film. It was
found that the MCA contribution $K_{\text{u}}^{\text{MCA}}$
to the UIP-MA constant $K_{\text{u}}$ was only slightly shifted
after step decoration and the shift was  much smaller than
experimental values (cf. Sec. \ref{sec-results}). With the same TB
model, shifts of $K_{\text{u}}^{\text{MCA}}$ by similar or smaller magnitude were
found for steps on Co films with thicknesses
other than 5 ML (in the range 2 up to 10 ML).

The present theoretical study focuses on explaining the mechanisms
by which decoration of Co steps with Cu atoms changes UIP-MA  in
vicinal Co/Cu(001) systems. We calculate
$K_{\text{u}}^{\text{MCA}}$  for stepped Co films using an
improved TB model. This model includes crystal field splittings
(CFS) introduced into the on-site energies of $d$-orbitals on
every atom in the system. Similar, though simplified, CFS have
previously been included and shown to play a crucial role in
calculations of the out-of-plane MA constant of flat films and
multilayers.\cite{bruno89,mc12,mc-p1,mc-p2,mc-p3,mc-2003} Here,
we also investigate how the UIP-MA constant
is modified by  the relaxation in the Co film structure around the step.
In addition, a potentially significant effect of
the charge transfer (CT) between Cu and Co atoms
in the decorated steps is studied.
We do not include the contribution
to  UIP-MA coming from the dipole-dipole interaction because it
remains almost unchanged when the Co steps are decorated with
non-magnetic atoms (cf. Sec. \ref{sec-results}, Ref.
\onlinecite{mc-mp-to-submit}).

\section{MODEL}
\label{sec-model}

\subsection{Geometry. Structure relaxation}
\label{subsec-geomstruc}
 In experiment, cobalt films deposited with molecular beam epitaxy
 grow pseudomorphically  on the vicinal (001) Cu surface
 for film thicknesses larger than 2 monolayers (ML) and less than
 10-15 ML.\cite{allenspach96c}. Thus, the Co film surface replicates the
underlying stepped substrate surface quite well in this thickness
regime. When submonolayer amount of copper is next deposited on
such vicinal films, Cu atoms decorate the Co step edges as has
been shown by scanning-tunneling-microscope
measurements.\cite{allenspach-sur-sci97}

\begin{figure}
\includegraphics*[width=\columnwidth]{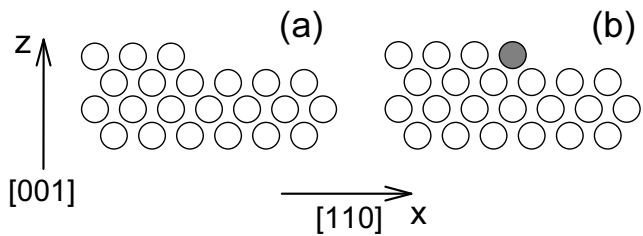}
\caption{Geometric structure of (001) fcc Co($N/N-1$) step:
(a) undecorated, (b) decorated with one Cu atomic wire
(shaded circle).}
\label{fig-step1}
\end{figure}

 The miscut angle $\alpha$ determines the terrace width
and consequently the number $N_{\text{rows}}$ of atomic
rows (lying parallel to $[1\overline{1}0]$) 
on each terrace: \cite{kawakami98a}
\begin{equation}
\label{nterrace} N_{\text{rows}}=\frac{1}{\sqrt{2}\tan{\alpha}}
\end{equation}
The steps on vicinal surfaces  are normally well separated.
Indeed, $N_{\text{rows}}$ is equal to approximately 400, 25, 12
atomic rows for the vicinal Co/Cu(001) systems with small miscut
angles $\alpha=0.1^{\circ}, 1.6^{\circ}, 3.4^{\circ}$,
respectively, which were studied experimentally in Refs.
\onlinecite{allenspach-apl97,allenspach-jap-review,
allenspach95a,allenspach95b,allenspach96a,allenspach96b,allenspach96c}.
 Therefore, as the decoration of a
given step is expected to change MCA only locally, we restrict our
theoretical model to a single monoatomic step on Co (001) fcc film.
For further simplicity, we assume that the step is present only on
the upper film surface and the bottom surface is flat
 so the system, hereafter called Co($N/N-1$) step, is actually built of
two joined semi-infinite slabs with thicknesses of $N$ and $N-1$ monolayers,
correspondingly [Fig. \ref{fig-step1}(a)].
For the decorated systems, there are one
or more Cu wires attached to the step edge on the upper surface on
the Co film [Fig. \ref{fig-step1}(b)].
For later reference, we assume that
the axes $x$, $y$, $z$ are along the directions [110],
$[1\overline{1}0]$ and [001], correspondingly, and the step-edge Co wire is
located
at the position $x=0$, $z=0$.

\begin{figure}
\includegraphics*[width=\columnwidth-2.0cm]{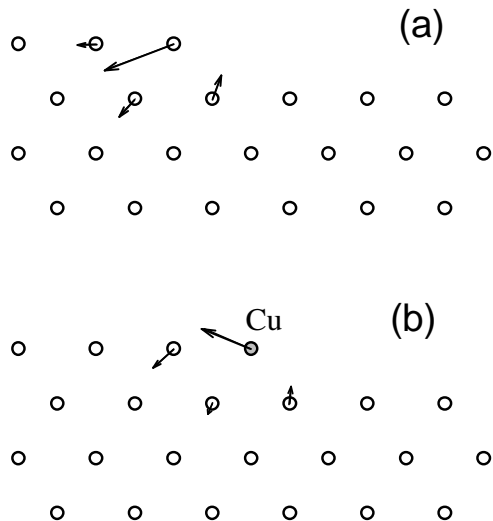}
\caption{Relative displacements (arrows) 
of the atomic positions in the biaxially strained lattice around
the edge of the $(001)$ fcc Co($4/3$) step: (a) undecorated, (b)
decorated with one Cu wire. The arrows are 30 times longer than
the actual displacements relative to the lattice. }
\label{fig-relax}
\end{figure}

Due 1.8\% mismatch between the lattice constants of cobalt and
copper, the cobalt film is subject to biaxial strain. This leads to
tetragonal distortion of the Co film lattice: the in-plane distances
between Co atoms are expanded, to match the Cu lattice constant,
while the distances between Co planes
become compressed in comparison to bulk fcc Co lattice.
In the present work, we consider Co($N$/$N-1$) steps
of thicknesses  $N\leq 10$ which have either
(i) perfect or (ii) tetragonally distorted fcc lattice.
In the latter case,  we allow for additional relaxation of the atomic
positions in the vicinity of the step.
The relaxed atomic positions  ${\bm R}_i$ are obtained by minimizing the system
energy
\begin{equation}
\label{eq-en-relax} E({\bm R}_1,{\bm R}_2,\ldots)= E_{\text B}({\bm
R}_1,{\bm R}_2,\ldots) + \frac{1}{2}\sum_{ij,i\neq j}
V_{\text{P}}({\bm R}_i-{\bm R}_j) .
\end{equation}
It includes the band energy $E_{\text B}$ and the pair potentials
$V_{\text{P}}$ which are given by the analytical formulae with
parameters fitted \cite{bazhanov00} to reproduce various elastic
properties either known from experiment for Cu or found with
first-principle methods for Co and Co/Cu systems. This
semi-empirical model, based on the second moment tight-binding
approximation, was  previously used to find strain relief and
shape evolution for a Co island on Cu
surface.\cite{bazhanov-co-island}
Here, we use the model to
calculate the displacements of atomic positions around the step
edge.
These displacements (magnified by a factor of 30) are
depicted for the Co(4/3) step in Fig. \ref{fig-relax}; very
similar shifts are found for Co(8/7) step. They are shown relative
to the biaxially strained slabs, Co($N)$ and Co($N-1$), for which
the interlayer spacings are also obtained with
Eq. (\ref{eq-en-relax}).

\subsection{Tight-binding model. Crystal field splittings}
\label{subsec-tb-model}

The electronic structure is described with an extended TB
hamiltonian
\begin{eqnarray}
\label{eq-ham-TB} H_{\text{TB}} =\sum_{i} \sum_{\mu\nu} \sum_{\sigma}
V_{i\mu\nu}^{\sigma}
c^{\dag}_{i\sigma\mu} c_{i\sigma\nu}+ \nonumber \\
{\sum_{ij}}'  \sum_{\mu\nu} \sum_{\sigma}T_{i\mu;j\nu}^{\sigma}
c^{\dag}_{i\sigma\mu} c_{j\sigma\nu} +H_{\text{SO}}(\theta,\phi).
\end{eqnarray}
Here, $\mu$, $\nu$ denote any of nine orbitals, of $s$, $p$, or $d$
symmetry, located on every atom (Co or Cu),  labeled with index $i$
or $j$; $\sigma$ is spin and $c^{\dagger}$, $c$ are creation and
annihilation operators, respectively. The hamiltonian depends on the
magnetization direction $(\theta,\phi)$ through the SO interaction
\begin{equation}
\label{hamil-so} H_{\text{SO}}(\theta,\phi)  =   \sum_{i}\sum_{\mu
\nu}\sum_{\sigma \sigma '} \xi_{i} \bra\mu \sigma |\bm{L} \cdot
\bm{S}|\nu \sigma '\ket c^{\dagger}_{i\mu \sigma} c_{i\nu \sigma '};
\end{equation}
where the  elements
$\bra\mu \sigma |\bm{L} \cdot \bm{S}|\nu \sigma '\ket $
are  the known analytical  functions of $\theta$,
$\phi$ (cf. Ref.\ \onlinecite{abate-asdente}).
We assume the SO
constant $\xi_{i}=0.085$ eV for Co atoms\cite{mc-p1,mc-p2,mc-p3},
and  $\xi_{i}=0.10$ eV for Cu atoms;\cite{mackintosh80,lorenz96,SO-p-orbitals}

The elements  $T_{i\mu,j\nu}^{\sigma}$ describing electron hopping
are calculated with the Slater-Koster (SK)
formulae;\cite{slater-koster} the sum over $j$ in Eq.
(\ref{eq-ham-TB}) includes the first and second nearest neighbors for every
atom $i$.  We assume the values of two-center hopping integrals
found for paramagnetic fcc bulk Co in Ref. \onlinecite{papa}  by
fitting to the {\it ab initio} band structure. When strain and
structure relaxation are present, the hopping elements
$T_{i\mu,j\nu}^{\sigma}$  are obtained by using the canonical
scaling law: $|{\bm R}_i-{\bm R}_j|^{-(l+l'+1)}$; here $l$ and $l'$
are the orbital quantum numbers for orbitals $\mu$ and $\nu$ centered
on atoms $i$ and $j$ which are located at the positions ${\bm R}_i$
and, ${\bm R}_j$, accordingly; cf. Ref. \onlinecite{pettifor}. The
hopping two-center integrals between Co and Cu atoms are
approximated by the geometric means of the corresponding Co and Cu
integrals.

The generalized on-site energies $V_{i\mu\nu}^{\sigma}$ (also known
as on-site potentials)
\begin{equation} \label{eq-gen-onsite}
V_{i\mu\nu}^{\sigma}= U_{i l;\sigma} \delta_{\mu\nu} +
V_{i\mu\nu}^{\text{CFS}}
\end{equation}
consist of two terms. The first one, diagonal in $\mu$, $\nu$, has
a common value $U_{i l;\sigma}$ for all orbitals $\mu$ of given
angular symmetry, $s$, $p$ or $d$ (corresponding to $l=0,1$ or
$2$). The other term describes CFS, i.e., possibly different
values of $V_{i\mu\mu}^{\sigma}$ for various orbitals $\mu$ with
the same $l$; in addition, non-zero elements
$V_{i\mu\nu}^{\text{CFS}}$ with $\mu\neq\nu$ may exist. The CFS
arise because  various orbitals centered on a given atom $i$ are
oriented differently with respect to neighboring atoms which
contribute to the electron potential $V^{\sigma}(\mathbf{r})$
on and around the atom $i$. The TB parameterizations for bulk cubic crystals,
like that in Ref. \onlinecite{papa}, include CFS, as the splitting
between the on-site energies of $t_{\text{2g}}$ and $e_{\text g}$
orbitals. However, CFS are often neglected in TB models for
systems with lower symmetry like surfaces (cf., e.g., Ref.
\onlinecite{dorantes-davila, llois02}) even though  the
corresponding CFS are larger than in bulk.\cite{galagher} The
splitting between the on-site energies of $d$ orbitals oriented
out-of-plane and those lying in-plane, located on atoms at the
film surface (or at an interface) has previously  been used  in
the MA calculations for thin films and
multilayers.\cite{bruno89,mc12,mc-p1,mc-p2,mc-p3,mc-2003} It
has been shown that such approximate CFS have a large effect on
the MA energies\cite{bruno89,mc12}

In the present paper, we calculate  $V_{i\mu\nu}^{\text{CFS}}$ for
$d$ orbitals $\mu$, $\nu$ centered on an atom $i$ as the sum
\cite{chadi89}
\begin{equation} \label{eq-vcfs}
V_{i\mu\nu}^{\text{CFS}}={\sum_j}' v_{i\mu\nu;j}^{\text{CFS}}
\end{equation}
over its first nearest neighbors; we neglect CFS for $s$ and $p$
orbitals and all  $sp$, $sd$, $pd$ off-diagonal terms
$V_{i\mu\nu}^{\text{CFS}}$. In Eq. (\ref{eq-vcfs}), the terms
$v_{i\mu\nu;j}^{\text{CFS}}$ are given by expressions formally
identical\cite{note-cfs-sk-sign}  to the SK formulae where the
two-center hopping integrals $V_{dd\sigma}$, $V_{dd\pi}$,
$V_{dd\delta}$ are replaced with similar parameters
$v_{dd\sigma}^{\text{CFS}}(m,m')$, $v_{dd\pi}^{\text{CFS}}(m,m')$,
$v_{dd\delta}^{\text{CFS}}(m,m')$ describing CFS. These two-center
CFS  parameters depend on the types $m$ and $m'$ of the atoms $i$
and $j$,  correspondingly, each of which is either Co or Cu.

This method of calculating CFS in TB models has been presented in
Ref. \onlinecite{chadi89} and used in some theoretical
papers.\cite{mercer94,cohen97,blackman01,blackman02} A more
general approach to the on-site energies is described in Ref.
\onlinecite{stiles97} while  a different method using the Fourier
transform of the  potential  is
applied to find the CFS in the bulk cubic crystals
in an early work on the TB model by Callaway and Edwards.
\cite{dme60}
The expressions for
$V_{i\mu\nu}^{\text{CFS}}$ and $v_{i\mu\nu;j}^{\text{CFS}}$ can be
derived similarly as the SK formulae, i.e., by approximating the
potential $V^{\sigma}(\mathbf{r})$ with the sum of spherically
symmetric atomic potentials
$v_j^{\sigma}(|\mathbf{r}-\mathbf{R}_j|)$ and expressing the
orbitals $\mu$ and $\nu$ centered on atom $i$ in terms of the
orbitals with the angular momentum quantized along the axis
joining the atoms $i$ and $j$. Then, the on-site energies
$V_{i\mu\nu}^{\sigma}$, found as the matrix elements of the
hamiltonian $\frac{1}{2}\nabla^2 + V^{\sigma}(\mathbf{r})$,
 include the terms $\langle i \mu|v_j^{\sigma}|i\nu\rangle$ which,
 after neglecting their spin-dependence, are
denoted as $v_{i\mu\nu;j}^{\text{CFS}}$ in Eq. (\ref{eq-vcfs}).
One concludes here that the elements $v_{i\mu\nu;j}^{\text{CFS}}$
are, in general, {\em not} symmetrical in  $i$, $j$, and nor do
the two-center CFS parameters, like
$v_{dd\sigma}^{\text{CFS}}(m,m')$,
need to have the same values for
$(m,m')=(\text{Co},\text{Cu})$ and $(m,m')=(\text{Cu},\text{Co})$.

Because, the term $V_{i\mu\nu}^{\text{CFS}}$ becomes
$V_{i\mu\nu}^{\text{CFS}}+c\delta_{\mu\nu}$ under the uniform shift
$v_{dd\eta}^{\text{CFS}}\rightarrow v_{dd\eta}^{\text{CFS}}+c$
($\eta=\sigma,\pi,\delta$),
 we can include $v_{dd\delta}^{\text{CFS}}$ into the diagonal on-site term
 $U_{i l;\sigma}$, Eq. \ref{eq-gen-onsite},
 and assume that $v_{dd\delta}^{\text{CFS}}=0$ hereafter.
We determine the values of $v_{dd\sigma}^{\text{CFS}}$ and
$v_{dd\pi}^{\text{CFS}}$  by fitting the energies
obtained with our tight-binding model for paramagnetic Co and Cu monolayers
to the {\it ab initio} bands.\cite{wimmer84}
As a result, we have found the values
$v_{dd\sigma}^{\text{CFS}}(m,m)=-0.17\;\text{eV}$ and
$v_{dd\pi}^{\text{CFS}}(m,m)=-0.10\;\text{eV}$, valid for both
$m=\text{Co}$ and $m=\text{Cu}$
within the accuracy of the fitting method.
The same values are assumed for the parameters $v_{dd\sigma}^{\text{CFS}}(m,m')$,
$v_{dd\pi}^{\text{CFS}}(m,m')$ used to find  the contribution of
a Cu atom to the on-site energy
$V_{i\mu\nu}^{\sigma}$ on a neighboring Co atom,
i.e., for $(m,m')=(\text{Co},\text{Cu})$,
or vice versa, i.e., for $(m,m')=(\text{Cu},\text{Co})$.
Thus, we can obtain, e.g., the splitting of 0.10 eV between
the on-site energies of $zx$ and $yz$ orbitals  for atoms
located on the step edge. Similarly, we find that
the orbitals $3z^2-r^2$ and $x^2-y^2$ are split by 0.20 eV at the
flat film surface; this value is close to the splitting of 0.22 eV
between out-of-plane and in-plane orbitals assumed previously
for Co slabs.\cite{mc12,mc-2003}
We do not scale the two-center CFS parameters  when the inter-atomic distances
change slightly in systems with relaxed structure;  the canonical scaling
law used for the hopping integrals seems not to be valid in this case.

The on-site energies  of down- and up-spin $d$ orbitals on Co atoms are
split by the exchange interaction:
\begin{subequations}\label{eq-v-ex-split}
\begin{eqnarray}
U_{il\downarrow}&=&U_{il}^{(0)}+\frac{1}{2} \Delta_{\text{ex}}^{(i)}, \\
U_{il\uparrow}&=&U_{il}^{(0)}-\frac{1}{2} \Delta_{\text{ex}}^{(i)}
\end{eqnarray}
\end{subequations}
($l=2$). The exchange splitting
\begin{equation}\label{eq-ex-split}
\Delta_{\text{ex}}^{(i)}= \Delta_{\text{ex}}^{\text{bulk}}
\frac{M_i}{M_{\text{bulk}}}
\end{equation}
on a Co atom $i$ is proportional to its spin moment $M_i$.
Here, we assume that the exchange splitting in bulk fcc Co
is $\Delta_{\text{ex}}^{\text{bulk}}=1.8\;\text{eV}$
(cf. Ref. \onlinecite{freeman88}) which, in the present TB model, leads to
the  bulk spin magnetic moment $M_{\text{bulk}}=1.57\;\mu_{\text{B}}$,
close to the {\em ab initio} values.\cite{freeman86,skriver92}
No exchange splitting is assumed for $s$ and $p$ orbitals on Co atoms and
for all orbitals on Cu atoms.

To find the yet undetermined on-site energy terms $U_{il}^{(0)}$
we require charge
neutrality, assumed separately for the $d$ orbitals,
\begin{equation}
\label{eq-ndcond} n_d^{(i)}=n_d^{\text{bulk};m} .
\end{equation}
and the $s$ and $p$ orbitals,\cite{note-sp-shift}
\begin{equation}
\label{eq-nspcond} n_{sp}^{(i)}=n_s^{(i)}+n_p^{(i)}=n_s^{\text{bulk};m} +
n_p^{\text{bulk};m} .
\end{equation}
 The quantities $n_s^{(i)}$, $n_p^{(i)}$, and
$n_d^{(i)}$ are the projected $s$, $p$, and $d$ occupations on an
atom $i$ while $n_s^{\text{bulk};m}$, $n_p^{\text{bulk};m}$, and
$n_d^{\text{bulk};m}$ are the corresponding values found
with the present TB model for the respective bulk metal $m$:
ferromagnetic fcc Co  or paramagnetic Cu.
The conditions (\ref{eq-ndcond}), (\ref{eq-nspcond})
are  well satisfied at transition-metal
surfaces as found in the {\em ab initio} calculations
\cite{skriver92} and have been used previously in TB models (cf.,
e.g., Refs. \onlinecite{falicov82,mc-au-surf-sci,mc-2003}).

According to Eqs. (\ref{eq-ndcond}), (\ref{eq-nspcond}),
we usually assume that there is {\em no} CT in the investigated systems.
However, for selected decorated Co($N/N-1$) steps,
we study the effect of CT
between Cu and Co atoms which leads to non-zero
charges $-q_i|e|$ on atoms $i$ around the step edge.
For this purpose,
it is assumed here that a Co atom
acquires $q_{\text{tr}}$ electrons from {\em each} of its
first-nearest Cu neighbors
so that  $-q_{\text{tr}}|e|$ is the charge transferred per one Cu-Co bond.
In consequence,
the  corresponding electron occupations $n_d^{(i)}$ and $n_{sp}^{(i)}$,
originally given by the right-hand sides of
Eqs. (\ref{eq-ndcond}), (\ref{eq-nspcond}),
have to be modified slightly.
So, we assume  that only the $d$-orbital occupation $n_d^{(i)}$ changes,
by the amount $q_i>0$, on Co atoms,
due to a relatively large value of the
local $d$-orbital-projected density
of states (DOS) at the Fermi level.
Further, it is assumed that the changes of the occupations
$n_d^{(i)}$ and $n_{sp}^{(i)}$ on Cu atoms are equal:
$\Delta n_d^{(i)}=\Delta n_{sp}^{(i)}=q_i/2$.
It should be noted here that the actual
distribution between the $sp$ and $d$ orbitals
of the electron deficiency $q_i=-5q_{\text{tr}}$
on the decorating Cu atoms
is found to be irrelevant as far as UIP-MA
is concerned.
The approximate value $q_{\text{tr}}=0.025$ is obtained by fitting,
within the applied TB model,
the decrease $M_i-M_{\text{bulk}}$  of the Co magnetic moment
$M_i$ at the Co/Cu interface in the (001) fcc Co(6)/Cu(10) slabs
to the  {\em ab initio} value $M_i-M_{\text{bulk}}=-0.06 \mu_{\text{B}}$
found in Ref. \onlinecite{skriver99}.

\subsection{Calculation methods: on-site energies, magnetocrystalline anisotropy}
\label{subsec-calc-methods}

The orbital-projected occupations $n_s^{(i)}$, $n_p^{(i)}$,
$n_d^{(i)}$, as well as the atomic moments $M_i$, can be obtained
easily once the diagonal part of the system's Green function (GF)
$G(z)=(z-H_{\text{TB}})^{-1}$ is known.\cite{mc-p1,haydock} We
determine the GF elements $\langle i\mu\sigma|G(z)|i\mu\sigma
\rangle$  for a number of atoms $i$ around the step by using  the
recursion method (RM).\cite{haydock}
We do this for all atomic wires which are distant by less
than 1.5 lattice  constants ($a_{0})$ in the $x$ ( i.e.,
$[1\overline{1}0]$) direction from the step edge;
in each wire,  we choose one atom with the smallest
$y$ (=0 or $a_{0}/(2\sqrt{2})$) coordinate. 
The on-site energy terms $U_{il}^{(0)}$ are then found for the
chosen atoms by solving Eqs. (\ref{eq-ndcond}), (\ref{eq-nspcond})
(appropriately modified for $q_{\text{tr}}\neq 0$) iteratively.
For every iteration, the new value of $U_{i2}^{(0)}$
for an atom $i$ (as well as for all other, equivalent, atoms
belonging to the same wire as the atom $i$) is expressed only with
the values of $U_{i2}^{(0)}$ and $n_d^{(i)}$ found in previous
iteration steps. This means that, for each atom $i$, Eq.
(\ref{eq-ndcond}) is solved as if it were a  non-linear equation
for a single variable $U_{i2}^{(0)}$. Similar assumption is made
for  Eq. (\ref{eq-nspcond}) and its dependence on $U_{i0}^{(0)}$
and $U_{i1}^{(0)}$. \cite{note-usp} Despite such approximate way
of solving the multi-atom set of Eqs. (\ref{eq-ndcond}),
(\ref{eq-nspcond}), which are obviously inter-independent to some
extent, only around 15 iterations are usually needed to fulfill
Eqs. (\ref{eq-ndcond}), (\ref{eq-nspcond}) for all atoms $i$ with
the  absolute error less than 0.0001. 
A similar technique was previously
successful for determination of layer on-site energies in thin magnetic
films and multilayers. \cite{mc12,mc-2003}
The success of such algorithm proves {\em a posteriori}
that the occupation $n_d^{(i)}$ on an atom $i$ depends {\em mainly}
on the term $U_{i2}^{(0)}$ (i.e., for $l=2$) determining the  on-site energies
of $d$-orbitals  on the same atom $i$;
similarly, $n_s^{(i)}+n_p^{(i)}$ depends mainly on $U_{i0}^{(0)}$
and  $U_{i1}^{(0)}$ ($l=0,1$).

The occupations  $n_s^{(i)}$, $n_p^{(i)}$, $n_d^{(i)}$ and the
moments $M_i$  for atoms $i$, which lie farther than $1.5a_{0}$
from the step edge, are to a good approximation not affected by
its presence. Therefore, the on-site energy terms $U_{il}^{(0)}$
for these atoms are the same as for flat slabs,\cite{mc-2003} $N$
or $N-1$  monolayers thick, and they are determined separately,
prior to the main calculation for the Co($N/N-1$) step.

The magnetocrystalline  contribution
to the UIP-MA  constant  $K_{\text{u}}$, Eq. (\ref{Ku-dif-e}), is calculated
with use of the force theorem\cite{daalderop90,weinberger96}:
\begin{equation}
\label{eq-ku-mca}
K_{\text{u}}^{\text{MCA}} =
\Omega(\theta,\phi=\frac{\pi}{2})-\Omega(\theta,\phi=0)  \;\;\;\;
(\theta=\frac{\pi}{2}).
\end{equation}
The electronic grand thermodynamic potential for
a given magnetization direction $(\theta,\phi)$,
\begin{equation}
\label{eq-grand-pot}
\Omega(\theta,\phi) = -k_{\text{B}} T \int dE \;
\ln\left [ 1 + \exp \frac{\epsilon_{\text{F}}-E}{k_{\text{B}} T} \right ]  n(E) ,
\end{equation}
is expressed in terms of the GF via  the total density of states
(DOS)
\begin{equation}
\label{eq-dos}
n(E) = -\frac{1}{\pi} \text{Im}\; \text{Tr}\; G(E+i\delta)
\end{equation}
($\delta \rightarrow 0+$);
$\epsilon_{\text{F}}$ is the Fermi energy and
$k_{\text{B}}$ the Boltzmann constant (in Eqs. (\ref{eq-dos}),
({\ref{eq-dos-gdiag})
exclusively, $i$ denotes $\sqrt{-1}$).
We use thermodynamic potential  $\Omega$ for a finite temperature $T$ (=300 K)
instead of zero-temperature energy $E$
because  such approach has previously been found to help
MA energies converge  for
flat thin films due to the thermal smearing
of the Fermi level.\cite{mc-p1,mc-2003}
However, it should be stressed that finite $T$ is used {\em only}
as  a convenient tool in the numerical calculations.
In particular, the present model
does not account for the decrease of the spontaneous magnetization
which occurs in real systems at finite temperatures due magnetic excitations
(cf. Refs. \onlinecite{mc-p1,mc-2003}).

We calculate the GF and hence the uniaxial in-plane
magnetocrystalline anisotropy (UIP-MCA) constant using two
distinct methods. The first one applies the standard RM  to find
the diagonal GF elements 
entering the expression for the DOS:
\begin{equation}
\label{eq-dos-gdiag}
n(E)= -\frac{1}{\pi} \sum_{j\kappa}  \text{Im}\;
\langle j\kappa | G(E+i\delta) | j\kappa \rangle .
\end{equation}
which comes from the general formula (\ref{eq-dos}) when
the trace is calculated with the states
\begin{equation}
| j\kappa \rangle =
\frac{1}{\sqrt{2}} \left( | j\mu\uparrow\rangle \pm
| j\mu\downarrow\rangle\right) .
\end{equation}
These states are chosen instead of the states $|
j\mu\uparrow\rangle$, $(| j\mu\downarrow\rangle$, because the
continued fractions that express the diagonal GF elements in
the RM converge much quicker in this case (cf. Ref.
\onlinecite{pick94}). However, still as many as 400 levels of
the continued fraction for $d$ orbitals and 120 for $s$ and $p$
ones are required to
calculate $K_{\text{u}}^{\text{MCA}}$ with the accuracy of 0.01
meV/(step atom) (i.e., per one atom on the step edge);
\cite{note-CF-for-d-only} similar number of
levels have been applied in Ref. \onlinecite{pick-bruno03}.

The UIP-MCA constant determined with RM is given by the sum of the atomic
contributions
\begin{equation}
\label{eq-ku-atomic-contrib}
K_{\text{u}}^{\text{MCA}}=\sum_j k_{\text{u;MCA}}^{(j)}.
\end{equation}
To reach the accuracy 0.01 meV/(step atom) for
$K_{\text{u}}^{\text{MCA}}$, it is sufficient to include the
contributions $k_{\text{u;MCA}}^{(j)}$ from all atomic wires which
are within the 3.5 $a_0$ distance from the step edge (cf. Figs.
\ref{fig-ku-ctrb-0cfs-0rlx}, \ref{fig-ku-ctrb-cfs-0rlx},
\ref{fig-ku-ctrb-cfs-rlx} and the discussion below); again we
choose one atom with the smallest $y$ coordinate in each atomic
wire.

A many-fold reduction of the computational cost in the RM
calculations of $K_{\text{u}}^{\text{MCA}}$, as well as
$n_s^{(i)}$, $n_p^{(i)}$, $n_d^{(i)}$ is obtained by truncating
the Co($N-1/N$) step, nominally infinite in-plane, to a finite
cluster with the square base which is centered at $x=y=0$ and has
its two sides  parallel to the step edge. It has been found that
the  effect of the introduced cluster boundaries is so small that
$K_{\text{u}}^{\text{MCA}}$ does not change by more than 0.01 meV/(step atom)
provided that the cluster size in the $x$  and $y$ directions
exceeds 50 $a_0$. This theoretical finding is interesting because
the coefficients in the continued fraction are obtained through
iterative application of the TB hamiltonian $H_{\text{TB}}$, Eq.
(\ref{eq-ham-TB}), on an initial state, like $|j\kappa \rangle$,
located on an atom close to the cluster center. Thus, these
coefficients start to be disturbed by the presence of the
boundaries already after around 25 iterations for the assumed
cluster size. Similarly, a cluster length and width of more than
30 $a_0$ guarantees that the errors in the occupations
$n_s^{(i)}$, $n_p^{(i)}$, $n_d^{(i)}$ and the moments $M_i$  are
less than 0.0001 for atoms lying close to the cluster center. This
being true despite the fact that the  number
of levels in the continued fraction
required for such high accuracy is 30 for $s$ and $p$
orbitals, and 120 for $d$ ones.

Because, a few hundred levels of the continued fraction
are  needed to obtain well-converged results for the UIP-MCA constant, this calculation
 becomes  difficult for Co($N/N-1$) steps with thicknesses
$N$ larger than 6 ML due to the long computation times
required. To remedy this situation, a new method, based on the
Mobius transformation (MT) approach \cite{umerski-mt} to calculate
the GF, has been developed. First, the  energy $E$ (or grand
potential $\Omega$) is shown to have the following general form:
\begin{equation}
\label{eq-e-mt}
E(\theta,\phi)=E_{\text{const}}+E_{\text{bulk}} L +E_{\text{osc}}(L).
\end{equation}
where $L$ is the length of the Co($N/N-1$) step after truncating
it only in the $x$ direction; the truncated system remains
infinite in the $y$ direction. The MT method allows us to
determine each energy term, present in Eq. (\ref{eq-e-mt}),
separately.
 The term $E_{\text{osc}}(L)$,
 is an oscillatory function of $L$ and comes from the
interference between the boundaries of the truncated step.
 This goes
to 0 for the infinite step, i.e., for $L\rightarrow \infty $. The
bulk term $E_{\text{bulk}}(\theta,\phi)$ is the sum of the
energies of the disconnected {\em flat} slabs, $N$- and
($N-1$)-monolayers thick, and as such it contributes only to the
cubic MA constant $K_1$ in the in-plane dependence of the step
energy on the magnetization direction $(\theta,\phi)$, Eq.
(\ref{MA-energy-gen}). The constant  term $E_{\text{const}}$
includes  the contributions $E_{\text{ends}}$ coming the
 boundaries of the truncated step (which are located at $x=-L/2$ and $x=L/2$).
These, unwanted contributions can be found separately
(by considering truncated flat slabs), and thus eliminated
from $E_{\text{const}}$,  leaving the the contribution
$E_{\text{step}}$ coming the step region alone:
\begin{equation}
\label{eq-mt-estep}
E_{\text{step}}=E_{\text{const}}-E_{\text{ends}}.
\end{equation}
Finally, the UIP-MCA constant is calculated as
\begin{equation}
\label{eq-mt-ku-by-estep}
K_{\text{u}}^{\text{MCA}}=
E_{\text{step}}(\phi=\frac{\pi}{2}) -
E_{\text{step}}(\phi=0)
\end{equation}
$(\theta=\frac{\pi}{2})$.
More details of this MT approach can be found in
Ref. \onlinecite{mc-au-surf-sci} while the full description
will be published elsewhere.\cite{umerski-future1}
The MT method is much more efficient computationally than RM
and allowed us to obtained accurate values of the UIP-MCA constants
for steps Co($N/N-1$) with thicknesses $N$ up to 10 ML.

\section{RESULTS. COMPARISON WITH EXPERIMENT}
\label{sec-results}

\begin{figure}
\includegraphics*[width=\columnwidth-1.0cm]{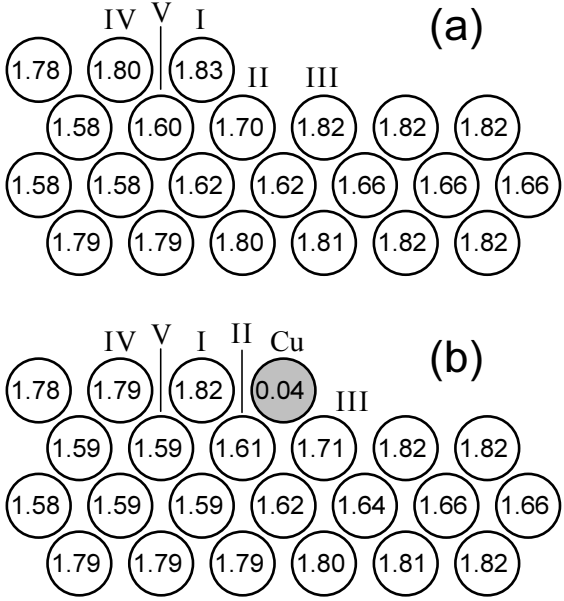}
\caption{Atomic magnetic moments $M_i$ (in $\mu_{\text{B}}$) for
the $(001)$ fcc Co($4/3$) step: (a) undecorated, (b) decorated
with one Cu wire. The results are obtained within the TB model
including {\em neither} CFS  {\em nor} structure relaxation (cf.
Sec. \ref{subsec-tb-model}). Selected Co atoms are marked with
Roman numbers for  reference in text.} \label{fig-mm-0csrlx0}
\end{figure}

\begin{figure}
\includegraphics*[width=\columnwidth-0.5cm]{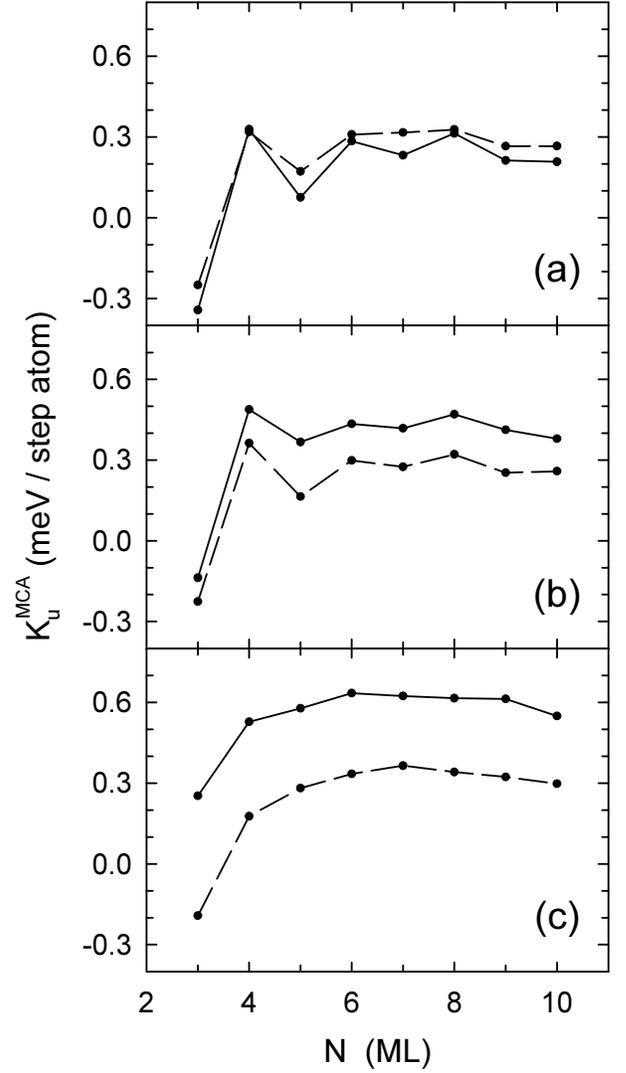}
\caption{UIP-MCA constant  $K_{\text{u}}^{\text{MCA}}$ [cf.\ Eq.\
(\ref{eq-ku-mca})] vs Co film  thickness $N$ calculated for
 $(001)$ fcc Co($N/N-1$) steps: undecorated (solid line)
and decorated with one Cu wire (broken line); cf. Fig. 1. The
results are obtained within the TB model (cf. Sec.
\ref{subsec-tb-model}) which includes: (a) {\em neither} CFS {\em
nor} structure relaxation, (b) CFS but {\em no} structure
relaxation, (c) {\em both} CFS and structure relaxation.}
\label{fig-ku-all}
\end{figure}

The atomic spin  magnetic moments obtained within the described TB
model by using RM are presented in Fig.  \ref{fig-mm-0csrlx0}. The
moments at the Co film surface are enhanced, in comparison to
bulk, and similar results have been found in {\em ab initio}
calculations. \cite{freeman88,skriver92} The moment of Co atoms at
the very step edge is  further enhanced. This confirms the general
rule that the local magnetic moment, as well as the local exchange
splitting, are the larger the smaller the atom's coordination
number, which is a consequence of
the reduced local $d$-orbital-projected density of states
due to the decreased hopping.
A closer comparison of Figs. \ref{fig-mm-0csrlx0}(a) and
\ref{fig-mm-0csrlx0}(b) 
reveals that the decorating Cu atom modifies the moments of the
neighboring Co atoms, due to the Co-Cu hybridization, as if it
were a Co atom; at the same time the Cu atom acquires only a very
small, induced, magnetic moment. In particular, the moments of Co
atoms labeled II and III in Fig. \ref{fig-mm-0csrlx0} become,
after decoration, very close to the moments of the atoms V and II,
respectively, before decoration.
This resembles the situation  at the interface
of Co and Cu films where the Co moments are very close to the bulk
Co moments.\cite{skriver99,blackman02} Similar behavior and values
of the moments  around the step edge have been previously found
with tight-binding linear-muffin-tin-orbital method in Ref.
\onlinecite{smirnov96-7}. Presently, we have also found that the
inclusion of CFS, biaxial strain and local structure relaxation,
discussed above, modifies magnetization very weakly: atomic
moments change  by no more than $0.02\; \mu_{\text{B}}$ in
comparison to those shown in Fig. \ref{fig-mm-0csrlx0}.

Slightly larger changes of Co atomic moments are induced by
the CT between Cu and Co atoms if we assume
that  it takes place.
Indeed,
we find for the decorated Co(4/3) step
that the moments on Co atoms I, II, III
decrease from $1.821\; \mu_{\text{B}}$, $1.599\; \mu_{\text{B}}$,
$1.722\; \mu_{\text{B}}$, respectively, for $q_{\text{tr}}=0$,
to the corresponding values
  $1.794\; \mu_{\text{B}}$, $1.544\; \mu_{\text{B}}$,
$1.668\; \mu_{\text{B}}$ found with the fitted value
$q_{\text{tr}}=0.025$ of CT per Cu-Co bond; all these moments are
obtained within the TB model with CFS
and structure relaxation included.
The moment change $\Delta M_i$ induced by  CT is very close
to the negative value of the local electron excess $q_i$
on Co atom $i$, this being $q_i=q_{\text{tr}}$ for atom I, and
$q_i=2q_{\text{tr}}$ for atoms II, III.
This happens because
the condition $\Delta n_d^{(i)}=q_i$ has been assumed to hold for  Co atoms,
on which the minority-spin component of DOS at the Fermi level is much
larger than  the majority-spin one.

Because, the decoration of Co($N/N-1)$ step with Cu atoms modifies
the Co magnetic moments only in the immediate vicinity of the Co
step edge (Fig. \ref{fig-mm-0csrlx0}), the resulting change in the
dipole-dipole energy is minute.\cite{mc-mp-to-submit} Thus, we can
neglect the dipolar contribution $K_{\text{u}}^{\text{dip}}$ to
the UIP-MA constant $K_u$ in the {\em present} study which is
focussed on the effect of the step decoration. However, in
general, this contribution is not negligible: it is positive and
favors a magnetization lying parallel to the step edge. The
calculations of $K_{\text{u}}^{\text{dip}}$ for magnetic films on vicinal surfaces
and networks of magnetic stripes deposited on flat or vicinal surfaces will
be presented elsewhere.\cite{mc-mp-to-submit}

 The calculations of the  UIP-MCA constant $K_{\text{u}}^{\text{MCA}}$
have been done in three consecutive stages. The CFS are included
in the TB model in the second stage and structure relaxations in
the third, while the initial TB model, i.e., in the first stage,
includes neither of the two above ingredients.
However, at every
stage, the on-site energy terms $U_{il}$ are found as described in
Sec. \ref{sec-model}. All the results for
$K_{\text{u}}^{\text{MCA}}$ are gathered in Fig.\
\ref{fig-ku-all}.
 In each of the three variants of the TB model,
the obtained $K_{\text{u}}^{\text{MCA}}$ is positive for both
decorated and undecorated Co($N/N-1$) steps with thicknesses $N\ge
4$. When no CFS nor structure relaxation are present the step
decoration with Cu has only small effect on
$K_{\text{u}}^{\text{MCA}}$: it is changed by no more than 0.05
meV/(step atom) and the shift is positive.
\cite{note-previous-work}
 The presence of CFS shifts the UIP-MCA constant $K_{\text{u}}^{\text{MCA}}$
 of the {\em undecorated} steps upwards by around 0.2 meV/(step atom);
 the inclusion of  structure relaxation results
in a further, slightly smaller, shift. At the same time,  the
average value of $K_{\text{u}}^{\text{MCA}}$ for decorated steps
remains almost unchanged, being close to 0.3 meV/(step atom) for
$N\ge 6$ ML, regardless of whether the CFS or structure relaxation
are included. As a net result, the  decoration decreases
$K_{\text{u}}^{\text{MCA}}$ by  $0.15-0.20$ meV/(step atom) or
$0.25-0.30$ meV/(step atom) when only CFS or both CFS and
relaxation, correspondingly, are included.

Another mechanism which, in principle, could affect the UIP-MCA constant
is CT between the decorating Cu
atoms and the neighboring Co atoms.
To study this possibility, we have  calculated the UIP-MCA constant
for the decorated Co(4/3) and Co(6/5) steps using the fitted value
$q_{\text{tr}}=0.025$ and its double, $q_{\text{tr}}=0.05$.
The results shown in Table \ref{table-ct} prove
that CT has very small effect on
the  UIP-MCA constant for $q_{\text{tr}}=0.025$.
For $q_{\text{tr}}=0.05$, the change of $K_{\text{u}}^{\text{MCA}}$
is  somewhat larger but it is still much smaller than the shift of
$K_{\text{u}}^{\text{MCA}}$ due to CFS or structure relaxation.
Thus, we conclude that CT is not  responsible for the UIP-MCA  shift
induced by the step decoration and it can be neglected in
the further discussion.
However, this conclusion should not be generalized to other systems
where larger interatomic charge transfers can take place and lead to more significant changes of MCA energies as a result of larger shifts of the local DOS with respect to the Fermi  level.
\begin{table}
\caption{UIP-MCA constant  $K_{\text{u}}^{\text{MCA}}$
[in meV/(step atom)]  of
 $(001)$ fcc Co($4/3$) and Co($6/5$)  steps decorated with one Cu wire.
 The results obtained within the TB model which includes CFS, structure relaxation, and CT of $q_{\text{tr}}$ electrons
 per Cu-Co bond.
\label{table-ct}  }
\begin{ruledtabular}
\begin{tabular}{c|c|c|c}
   &$q_{\text{tr}}=0$ & $q_{\text{tr}}=0.025$ & $q_{\text{tr}}=0.05$ \\
\hline
Co(4/3) &0.18 & 0.17 & 0.15 \\
Co(6/5) &0.33 & 0.32 & 0.28 \\
\end{tabular}
\end{ruledtabular}
\end{table}

The obtained $K_{\text{u}}^{\text{MCA}}$, shown in Fig. \ref{fig-ku-all},
oscillates with a period
close to 2 ML when  the step thickness $N$ increases. Similar 2-ML
oscillations of out-of-plane MCA constant have previously been
found for flat Co films \cite{mc-2003} and Co/Cu
multilayers.\cite{szunyogh97} It has been shown\cite{mc-2003} that
these oscillations originate from the quantum well (QW) states
present in the Co layer which have $\bm k$-vectors in the center
of the two-dimensional Brillouin zone and energies  close to the
Fermi level. Similar QW states are likely to be responsible for
the oscillations of $K_{\text{u}}^{\text{MCA}}(N)$ in the
Co($N/N-1$) steps. Interestingly, such oscillations are absent
once the atomic structure is allowed to relax (cf. Fig.
\ref{fig-ku-all}(c)); explanation of this behavior needs further
study, especially, in connection to the absence of such
oscillations in experiment. \cite{allenspach96a,allenspach96c}

To compare the obtained results with the experimental data found for
Co films on vicinal (001) Cu surfaces, we convert the measured
$H_{\text{s}}$ field, given
in kA/m, into the UIP-MA constant $K_{\text{u}}$:
\begin{equation}
\label{eq-hs-to-ku}
 K_{\text{u}}=1.254 \cdot 10^{-4} N N_{\text{rows}} H_{\text{s}} \;\;
 \text{[meV/(step atom)]} ;
\end{equation}
here $N_{\text{rows}}$ is the terrace width which depends on the miscut angle
$\alpha$ through Eq. (\ref{nterrace}).
In Ref. \onlinecite{allenspach96b}, the $H_{\text{s}}$ field is found
experimentally for $\alpha=3.4^{\circ}$ and the Co film thicknesses
$N=9.4$, 10.8, and 13.2 ML.
By taking the corresponding values,
$\Delta H_{\text{s}}=-13.5$, $-11.5$, and $-12.5$ kA/m, of  the  differences
$\Delta H_{\text{s}}=H_{\text{s}}^{\text{dec}}-H_{\text{s}}^{\text{undec}}$
 between the minimum $H_{\text{s}}$ field
found for the decorated films, $H_{\text{s}}^{\text{dec}}$, and
the field $H_{\text{s}}^{\text{undec}}$ for the undecorated film,
we obtain with Eq. (\ref{eq-hs-to-ku}) the respective experimental shifts
$\Delta K_{\text{u}}=-0.19$, $-0.19$, and $-0.25$ meV/(step atom)
of the UIP-MA constant.
A very similar value of $\Delta K_{\text{u}}$ is found from the data reported
in Ref. \onlinecite{kawakami98a} for 8-ML Co film deposited on slightly curved
Cu surface which has a variable miscut angle $\alpha$ $(\le 5.4^{\circ})$.
It was shown there that $H_{\text{s}}$ depends linearly on
 $\alpha$ and it vanishes
after deposition of an amount of Cu sufficient to decorate 70\% of
Co steps with single Cu wires. This result allows us to write down
the equation:
\begin{equation}
\label{eq-hs-kawakami}
0.3H_{\text{s}}^{\text{undec}}+0.7 H_{\text{s}}^{\text{dec}}= 0
\end{equation}
from which we find that $H_{\text{s}}^{\text{dec}}=-0.43
H_{\text{s}}^{\text{undec}}$ and $\Delta H_{\text{s}}=-1.43
H_{\text{s}}^{\text{undec}}$ . Thus, from the experimental value
\cite{kawakami98a} $H_{\text{s}}^{\text{undec}}\approx 150\;
\text{Oe}=11.94 \; \text{kA/m}$ found for $\alpha=3^{\circ}$,
we get $\Delta K_{\text{u}}=-0.23$
meV/(step atom). The experimental values of the UIP-MA shift
$\Delta K_{\text{u}}$ derived above are close to our theoretical
results obtained both with or without the structure relaxation,
provided that CFS are included. We regard the  remaining
discrepancy of 0.05-0.01 meV/(step atom) as very low since
calculated MA energies, even those found with {\em ab initio} methods
rarely match the experimental data perfectly. Good theoretical
prediction for $\Delta K_{\text{u}}$ confirms the  conclusion of
the experimental work reported in  Ref. \onlinecite{kawakami98a}
that the decrease of $K_{\text{u}}$ is `caused only by the Cu
adsorbates located near the step edges'. However, we do not obtain
the sign change of  $K_{\text{u}}$ after decoration which is usually
observed experimentally
\cite{allenspach95a,allenspach95b,allenspach96b,kawakami98a}
and leads to switching of the magnetization direction within the
plane. Indeed, after the decoration, $K_{\text{u}}^{\text{MCA}}$
remains positive (for $N\ge 4$); adding the dipolar contribution
$K_{\text{u}}^{\text{dip}}$ will push the resulting $K_{\text{u}}$
even higher. This implies that the theoretical energies
$K_{\text{u}}^{\text{MCA}}$ are too large in comparison with
experiment, both for undecorated and decorated steps. This
discrepancy is presumably due to some unaccounted terms of volume
type which are clearly present in the experimental
$H_{\text{s}}(N)$ dependencies.
\cite{allenspach-apl97,allenspach-jap-review} The origin of such
terms is not clear \cite{kawakami98a,kawakami99} and needs further
investigation. It should be noted here that the tetragonal
distortion of the biaxially strained fcc Co film lattice, which is
included in our calculations, retains the films fourfold in-plane
symmetry and, thus, gives no direct volume contribution to
UIP-MA.\cite{kawakami98a,kawakami99}

\begin{figure}
\includegraphics*[width=\columnwidth-0.5cm]{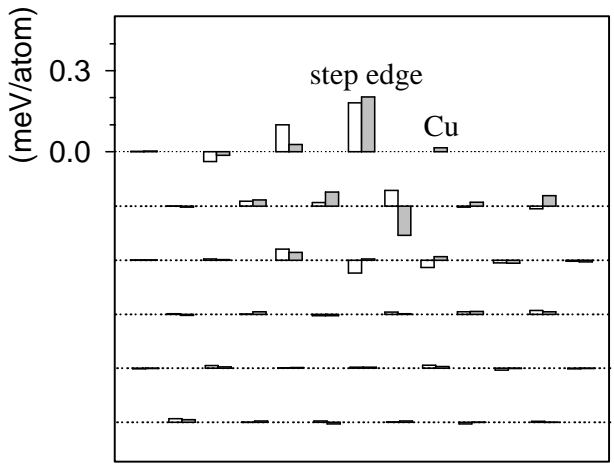}
\caption{Atomic contributions $k_{\text{u;MCA}}^{(i)}$ (bars)
to UIP-MCA constant  $K_{\text{u}}^{\text{MCA}}$
[cf.\ Eq.\ (\ref{eq-ku-atomic-contrib})]  for $(001)$ fcc Co($6/5$) step:
undecorated (white bars) and  decorated (grey bars) with one Cu wire.
The horizontal dotted lines mark (001) atomic planes;
atomic wires (Co or Cu) are located
at points where the bars touch the lines
which  also serve as the zero reference levels
for the bars.
The plotted data are obtained within the TB model
which includes {\em neither} CFS {\em nor} structure relaxation
(cf. Sec. \ref{subsec-tb-model}).}
\label{fig-ku-ctrb-0cfs-0rlx}
\end{figure}

\begin{figure}
\includegraphics*[width=\columnwidth-0.5cm]{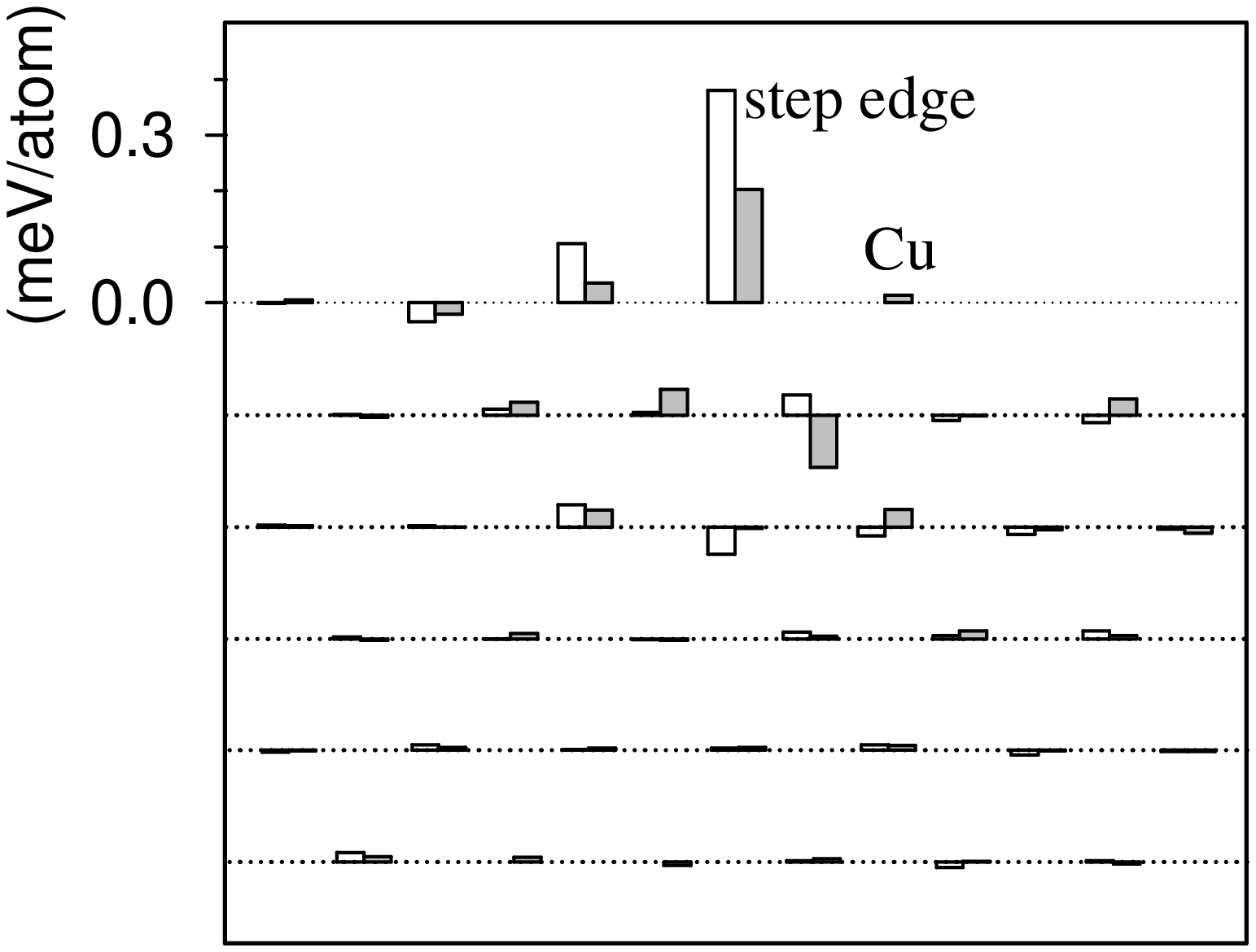}
\caption{Atomic UIP-MCA contributions $k_{\text{u;MCA}}^{(i)}$
 obtained for  $(001)$ fcc Co($4/3$) step within the TB model
which includes  CFS but {\em no} structure relaxation
(cf. Sec. \ref{subsec-tb-model}).
Other details as in Fig. \ref{fig-ku-ctrb-0cfs-0rlx} }
\label{fig-ku-ctrb-cfs-0rlx}
\end{figure}

\begin{figure}
\includegraphics*[width=\columnwidth-0.5cm]{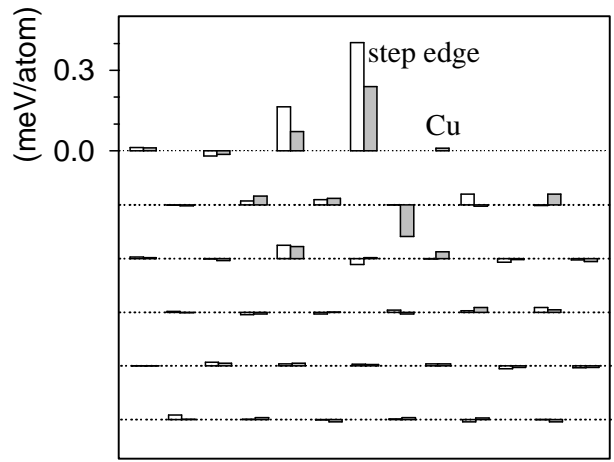}
\caption{Atomic UIP-MCA contributions $k_{\text{u;MCA}}^{(i)}$
 obtained for  $(001)$ fcc Co($4/3$) step within the TB model
which includes {\em both} CFS and structure relaxation
(cf. Sec. \ref{subsec-tb-model}).
Other details as in Fig. \ref{fig-ku-ctrb-0cfs-0rlx}}
\label{fig-ku-ctrb-cfs-rlx}
\end{figure}

To develop a better understanding the origin of the obtained shift
$\Delta K_{\text{u}}^{\text{MCA}}$, we examine the contributions
$k_{\text{u;MCA}}^{(i)}$ to the UIP-MCA constant from individual
atoms around the step; see Figs. \ref{fig-ku-ctrb-0cfs-0rlx},
\ref{fig-ku-ctrb-cfs-0rlx}, \ref{fig-ku-ctrb-cfs-rlx}. The largest
contribution $k_{\text{u;MCA}}^{(i)}$ comes from the Co atomic
wire at the very step edge, marked as atom I in Fig.
\ref{fig-mm-0csrlx0}. Other atomic wires that lie within the
distance $d_{w}$ of 1.5 $a_0$ from the step edge, give smaller,
but still quite sizeable contributions $k_{\text{u;MCA}}^{(i)}$.
The contributions from farther lying atomic wires decay very
quickly with increasing $d_{w}$. However, as already mentioned
above, the contributions from several tens of atomic wires with $0
\leq d_{w} \leq 3.5 a_{0}$ are required to obtain
$K_{\text{u}}^{\text{MCA}}$ with the accuracy of 0.01 meV/(step atom). The
step decoration with Cu atoms affects UIP-MCA in three different
ways. First, the additional electron  hopping between the Cu atoms
and the neighboring Co atoms affects contributions around the step
edge, mainly on atoms II and IV, for which the respective
contributions $k_{\text{u;MCA}}^{(i)}$ are shifted downwards after
decoration. However, shifts $\Delta k_{\text{u;MCA}}^{(i)}$ of
both signs are found for several atoms around the step and, thus,
they tend to cancel out to a large extent which explains why the
resulting total shift  $\Delta K_{\text{u}}^{\text{MCA}}$, shown
in Fig. \ref{fig-ku-all}(a), is relatively small. The inclusion of
CFS leads to a large shift of $k_{\text{u;MCA}}^{(i)}$ on the
step-edge atom (atom I), for the undecorated steps, while the
contributions from other atoms remain almost unchanged. This means
that the UIP-MCA is affected only very locally by CFS. Indeed, the
effect of CFS on atom I is canceled once the step is decorated
with Cu atoms; cf. Fig. \ref{fig-ku-ctrb-cfs-0rlx}. This happens
because  the decorating Cu atoms remove, via Eq.~(\ref{eq-vcfs}),
the splitting between the energies of in-plane orbitals, like $yz$
and $zx$, located on atom I. On the other hand, the structural
relaxation has a more spatially extended effect on UIP-MCA: it
results in small changes of the contributions
$k_{\text{u;MCA}}^{(i)}$ for several atoms, both for undecorated
and decorated steps. This is presumably the indirect effect of the
tetragonal lattice distortion which changes the electronic
structure. Furthermore, the local change of the geometric structure
induced by the step decoration modifies the shifts $\Delta
k_{\text{u;MCA}}^{(i)}$ also only locally, mainly on atoms I to V
around the step. In each variant of the TB model, the total
UIP-MCA shift, $\Delta K_{\text{u}}^{\text{MCA}}$, can be found
[with 0.01 meV/(step atom) accuracy] by adding shifts of
$k_{\text{u;MCA}}^{(i)}$ from atoms with $0 \leq d_{w} \leq 2
a_{0}$; cf. Ref. \onlinecite{mc-au-surf-sci}. It should also be
noted that the contribution to $K_{\text{u}}^{\text{MCA}}$ from
the decorating Cu atom is negligible.

\begin{figure}
\includegraphics*[width=\columnwidth-0.5cm]{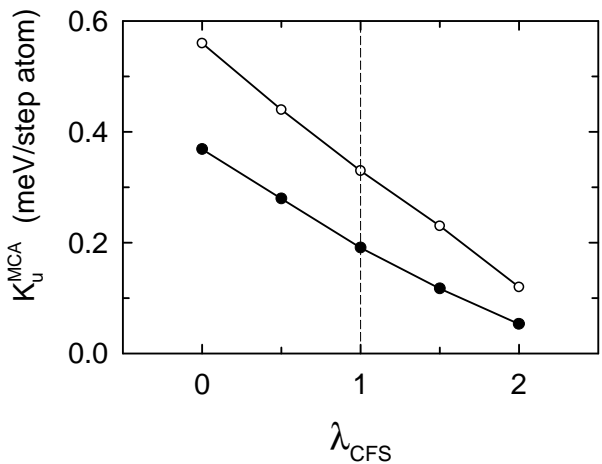}
\caption{UIP-MCA constant  $K_{\text{u}}^{\text{MCA}}$
for  the $(001)$ fcc Co($4/3$) (solid circles) and
 Co($6/5$) (open circles) steps decorated with
one Cu wire; cf. Fig. \ref{fig-step1}(b). The results are plotted
versus the multiplier $\lambda_{\text{CFS}}$ that modifies the CFS
contributions from the Cu atoms; cf. Eq. (\ref{eq-lam-cfs}); see
text. The results are obtained within the TB model including {\em
both} CFS and structure relaxation (cf. Sec.
\ref{subsec-tb-model}). The vertical dashed line denotes the
nominal value $\lambda_{\text{CFS}}=1$. } \label{fig-ku-cfscu}
\end{figure}

\begin{figure}
\includegraphics*[width=\columnwidth-0.5cm]{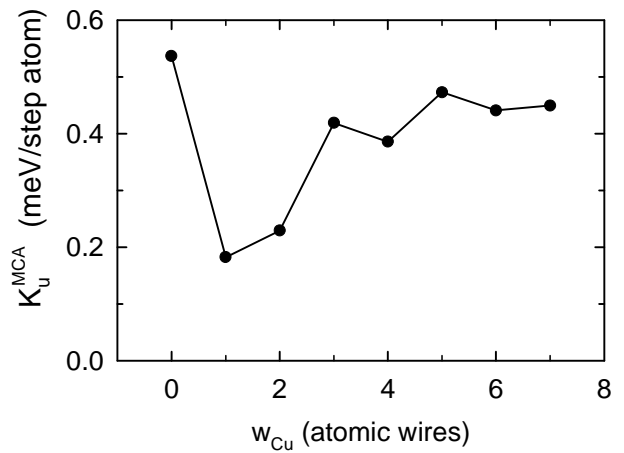}
\caption{UIP-MCA constant  $K_{\text{u}}^{\text{MCA}}$
for  the $(001)$ fcc Co($4/3$) step decorated with
$w_{\text{Cu}}$ Cu wires;
the data point for $w_{\text{Cu}}=0$ corresponds
to the undecorated step.
The results are obtained within the TB model
including {\em both} CFS and structure relaxation
(cf. Sec. \ref{subsec-tb-model}).}
\label{fig-ku-ncu}
\end{figure}

In order to better understand the mechanism by which CFS affect
UIP-MCA, we modify the contribution of the decorating Cu atoms to
CFS on the neighboring Co atoms. It is done by multiplying  the
CFS two-center integrals for $m=$Co, $m'=$Cu:
\begin{equation}
\label{eq-lam-cfs}
v_{dd\eta}^{\text{CFS}}(m,m') \rightarrow
\lambda_{\text{CFS}} \; v_{dd\eta}^{\text{CFS}}(m,m')
\end{equation}
with the common factor $\lambda_{\text{CFS}}$  for
$\eta=\sigma,\;\pi, \delta$; cf. Ref.
\onlinecite{note-cfs-scalling}.
In particular, for $\lambda_{\text{CFS}}=0$, when the Cu atoms
do not contribute to CFS on the Co atoms, we find that
the constant $K_{\text{u}}^{\text{MCA}}$ is larger by around 0.2 meV/(step atom)
than for the nominal value $\lambda_{\text{CFS}}=1$ when  the
decorated steps Co(4/3) and Co(6/5) with relaxed structure are considered;
cf. Figs.  \ref{fig-ku-cfscu}, \ref{fig-ku-all}(c).
In fact, the value of $K_{\text{u}}^{\text{MCA}}$ for $\lambda_{\text{CFS}}=0$
becomes closer to $K_{\text{u}}^{\text{MCA}}$ obtained for
the {\em undecorated} step in the TB model
including CFS ; the remaining difference  [0.08 meV/(step atom) for Co(6/5) step]
is due to the change in the local structure relaxation after
the step decoration.
We also note that with increasing $\lambda_{\text{CFS}}$, the UIP-MCA constant
of the decorated steps decreases, almost linearly,
however, it  remains positive for
$\lambda_{\text{CFS}}\lesssim 2$. Thus, we do not consider
possible inaccuracy in the derived two-center CFS parameters for Co and Cu
as the source of the lack of magnetization
switching in the present theoretical calculations.

Finally, we study how the UIP-MCA changes when more than one,
$w_{\text{Cu}}$, Cu wires are attached to the Co step edge,
in a similar way to that depicted in Fig.~\ref{fig-step1}.
The results
obtained for $0\leq w_{\text{Cu}}\leq 7 $ Cu wires, Fig.
\ref{fig-ku-ncu}, confirm the experimental finding
\cite{allenspach95a} that the maximum downward shift of
$K_{\text{u}}^{\text{MCA}}$ occurs at $w_{\text{Cu}}=1$. For
larger $w_{\text{Cu}}$, the value of
$K_{\text{u}}^{\text{MCA}}(w_{\text{Cu}})$ increases, showing
small oscillations, and saturates for
 $w_{\text{Cu}}\geq 6$ at a value below, though close to, the UIP-MCA constant
 $K_{\text{u}}^{\text{MCA}}$ for the undecorated step.
A similar tendency is also observed experimentally for Cu coverage
less than 1.2 ML (cf. Ref.
\onlinecite{allenspach95a,allenspach95b}) but the magnitude of
experimental UIP-MA shift, $\left|\Delta K_{\text{u}}\right|$,
is always larger than
$\frac{1}{2}\left| \Delta K_{\text{u}}(w_{\text{Cu}}=1)\right|$.

\section{CONCLUSIONS}
\label{sec-conclusions}

The present tight-binding calculations performed for a single step
on an ultrathin (001) fcc Co slab with in-plane magnetization show
that the magnetic anisotropy of vicinal Co films depends on
various factors. In particular, the inclusion of crystal field
splittings has been proved to be  {\em vital}  for the theoretical
results to reproduce well the experimentally observed decrease of
the uniaxial in-plane anisotropy energy induced by decorating the
Co steps with Cu atoms.  Thus, the generally assumed explanation that
the shift of the uniaxial anisotropy  energy is due to the local
hybridization between the Co and Cu atoms, is confirmed.
However,
this happens {\em not} through the extra available hopping between
Co and Cu atoms, but rather because the potential coming from
the decorating Cu atoms {\em removes}
the splitting which exists between the on-site energies of
$d$ orbitals, with different in-plane orientation,
on Co atoms at the edge of the undecorated step. 
The fact that the crystal field splittings
affect mainly the anisotropy contribution on the Co step-edge atom
points to the important role of horizontal, in-plane, bonds.
Similar conclusion
has previously been drawn from the experimental results through a
general analysis based on the  N\'{e}el's pair-bonding model.
\cite{kawakami98a}

The investigated shift of the uniaxial anisotropy
includes also a significant contribution from
the local relaxation of atomic structure around the step. This finding
is in agreement with the results of earlier studies
(cf., e.g., Ref. \onlinecite{baberschke97})
where  strain and structure relaxation in the {\em flat} films
have been shown to have great impact on magnetic anisotropy.

These conclusions remain valid also if the charge transfer between
Cu and Co atoms around the step edge is present since its effect on
the uniaxial anisotropy energy of the vicinal Co films has been shown
to be very weak.

\begin{acknowledgments}
We would like to acknowledge gratefully discussions
with Professor D.M. Edwards.
This work has been  supported financially
by the State Committee of Scientific Research (Poland)
under the research project no 2P03B 032 22 in the years 2002-2005.
\end{acknowledgments}


\begin{thebibliography}{99}
\bibitem{110-as-vicinal} A (110) surface of a fcc crystal
can be regarded as a vicinal surface with the  steepest miscut angle
possible ($\alpha=45^{\circ}$)  and the terrace width of just one atom.
\bibitem{ku-explanation} The value of  $K_{\text{u}}$
calculated with Eq. (\ref{Ku-Hs}) is given per unit volume of the
ferromagnetic film.
\bibitem{SO-p-orbitals} These values have been derived for  $d$ orbitals,
however
no seperate values are sought for $p$ orbitals as the SO
interaction between $p$ orbitals has been found to have negligigle
effect on MA; cf. Ref. \onlinecite{mc12}. This happens because both the exchange
splitting and the orbital-projected density of states  are much
smaller for $p$ orbitals than for $d$ ones. Also, note that the SO
interaction vanishes for $s$ states.
\bibitem{note-cfs-sk-sign} In general, when orbitals $\mu$ and $\nu$
have different orbital quantum numbers, $l$ and $l'$,
corrspondingly, the expressions for $v_{i\mu\nu;j}^{\text{CFS}}$
have an extra factor $(-1)^{(l+l')}$ in comparison with the SK
formulae.
\bibitem{note-sp-shift} The on-site energies of $s$ and $p$ orbitals
are subject to a common shift whose magnitute is chosen
to satisfy Eq. (\ref{eq-nspcond}).
\bibitem{note-usp}
In this case, the left-hand side of Eq. (\ref{eq-nspcond})
is treated to be the function of a common shift applied to $U_{i0}^{(0)}$,
$U_{i1}^{(0)}$
\bibitem{note-CF-for-d-only} We have also found that a smaller
number of CF levels is needed
for $d$ orbitals when the $spd$ hybridization is neglected.
\bibitem{note-cfs-scalling} Note that with $v_{dd\delta}^{\text{CFS}}(m,m')=0$
it is $v_{dd\eta}^{\text{CFS}}(m,m')-v_{dd\delta}^{\text{CFS}}(m,m')$
($\eta=\sigma,\;\pi$) that are really scaled.
\end{thebibliography}
\end{document}